\documentclass[aps,prb,twocolumn,amsmath,amssymb,superscriptaddress,showpacs]{revtex4-1}

\newcommand{\be}{\begin{equation}}
\newcommand{\ee}{\end{equation}}
\newcommand{\bea}{\begin{eqnarray}}
\newcommand{\eea}{\end{eqnarray}}
\usepackage[utf8]{inputenc}
\usepackage{amsfonts}
\usepackage{amsthm}
\usepackage{fontenc}
\usepackage{graphicx}
\usepackage{textcomp}
\usepackage{epstopdf}
\usepackage{braket}
\usepackage{mathtools}
\usepackage{dcolumn}
\usepackage{bm}
\usepackage{epsfig}
\usepackage[usenames,dvipsnames]{color}
\usepackage{booktabs}
\begin{document}
\newcommand{\abs}[1]{\lvert#1\rvert}
\title{Strained fold assisted transport in graphene systems}
\author{R. Carrillo-Bastos}
\affiliation{Facultad de Ciencias, Universidad Aut\'{o}noma de Baja California, Apartado Postal 1880, 22800 Ensenada, Baja California, M\'{e}xico.} \affiliation{Department of Physics and Astronomy, Ohio
University, Athens, Ohio 45701-2979, USA.}

\author{C. Le\'{o}n}
\affiliation{Instituto de F\'{i}sica, Universidade Federal Fluminense, Niter\'{o}i, Avenida
Litor\^{a}nea sn, 24210-340 Rio de Janeiro, Brasil.}

\author{D. Faria}
\affiliation{Instituto Polit\'{e}cnico, Universidade do Estado de Rio de Janeiro, Nova Friburgo, Rua Bonfim 25, 28625-570 Rio de Janeiro, Brasil.}

\author{A. Latg\'{e}}
\affiliation{Instituto de F\'{i}sica, Universidade Federal Fluminense, Niter\'{o}i, Avenida
Litor\^{a}nea sn, 24210-340 Rio de Janeiro, Brasil.}

\author{E. Y. Andrei}
\affiliation{Department of Physics and Astronomy, Rutgers University, Piscataway, New Jersey 08855, USA}

\author{N. Sandler}
\affiliation{Department of Physics and Astronomy, Ohio University, Athens, Ohio 45701-2979, USA.}

\begin{abstract}

\noindent Deformations in graphene systems are central elements in the novel field of {\it straintronics}. Various strain geometries have been proposed to produce specific properties but their experimental realization has been limited. Because strained folds can be engineered on graphene samples on appropriate substrates, we study their effects on graphene transport properties. We show the existence of an enhanced local density of states (LDOS) along the direction of the strained fold that originates from localization of higher energy states, and provides extra conductance channels at lower energies. In addition to exhibit sublattice symmetry breaking, these states are valley polarized, with quasi-ballistic properties in smooth disorder potentials. We confirmed that these results persist in the presence of strong edge disorder, making these geometries viable electronic waveguides. These findings could be tested in properly engineered experimental settings.
\end{abstract}
\maketitle 

\section{Introduction}
Since the original prediction by Fujita et al.\cite{Fujita}, edge states in pristine graphene have been heralded as ideal ballistic channels with potential in electronic applications. However, their detection has remained elusive due to their fragility in the absence of spin-orbit interactions (small in graphene)\cite{KaneMele}, and the presence of disorder at the edges. Numerical studies for ribbons with rough edge terminations confirm that edge disorder destroys ballistic motion along the edges\cite{Disorder}, and provides an explanation for the difficulties encountered in their experimental detection in transport measurements\cite{Kimribbons1,Kimribbons2,Klaus,GoldhaberGordon}. To better understand the nature of these states, Sasaki et al \cite{Sasaki} studied the effects of a highly localized strain defect along different crystal directions (modeled by a $\delta$-function gauge field in a Dirac model). The analytic solution showed the emergence of states along the zigzag direction with properties similar to those of edge states: pseudo-spin polarization, i.e., local sublattice symmetry breaking, and same flat band dispersion, but localized at the position of the deformation. These characteristics are well understood in terms of the effective pseudo-magnetic field generated by the deformation\cite{Moldovan, Peeters2, Peeters3, Carrillo, Martin, Blanter, Wakker, AntiPekka1, AntiPekka2}.

The suggestion of using strain to tailor electronic properties has been advanced by several authors \cite{GuineaGeim, Vitor1,Vitor2,Fernando,MFV,Barraza1,Moldovan,Prada1,Fogler,Vozmediano2,Falko1,Daiara,Gerardo1,Gradinar,Bahamon,Carrillo,Yang,Chan,Villegas,Neek-Amal1,Neek-Amal2,Polini}, and pursued in experimental settings\cite{Lau, Klimov, KimY, Crommie1, Jang}. Several groups have observed clear signatures of equilibrium properties in strained areas predicted by various models, such as pseudo-Landau levels and sublattice symmetry breaking in STM images\cite{Georgiou,Crommie2,Morgenstern,KimY}. Recent works have reported transport measurements on ribbon geometries\cite{Han-Kim,APLreview}, with one study revealing ballistic transport at room temperatures along nanoribbons deposited on terraced SiC substrates (thus subject to deformations)\cite{Baringhaus}. This particular geometry highlights the possibility of creating extended strained fold-like structures with unusual transport properties.
\begin{figure}
\includegraphics[scale=0.43]{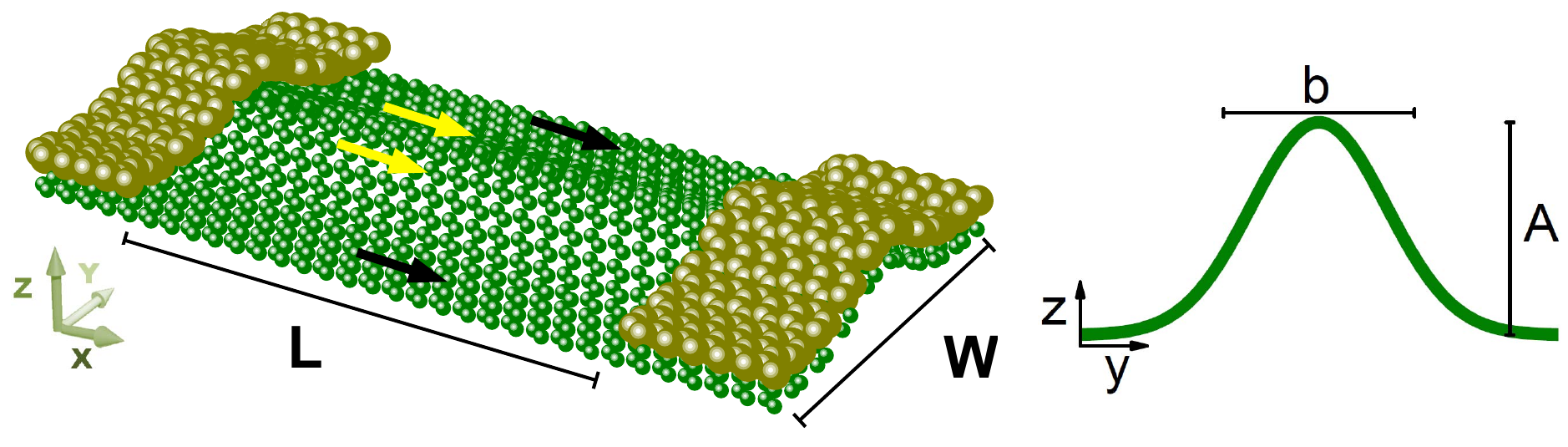}
\caption{(Color online) Schematic representation of deformed zigzag graphene nanoribbon
(width $W$ and length $L$) connected to leads with a Gaussian fold-like out-of plane
deformation (amplitude $A$ and width $b$). Arrows indicate valley currents as described in main text. Color code given in Fig. 4 (d). 
\label{Fig1}}
\end{figure}

While models for transport through strained areas have been the topic of several works, transport {\it along} deformed areas 
has been less explored. In fact, due to the peculiar properties of graphene electronic states under strain, extended deformed areas \cite{KimY}, may act as natural electronic waveguides. In this work we show that longitudinal out-of-plane deformations along a graphene membrane, generate extra conductance channels running parallel to the structure with the remarkable property of being valley polarized. As a consequence, a current injected parallel to the axis of the deformation will naturally split in space, with states from one valley running along the crest while states of the other valley run along the sides. These channels survive  in the presence of highly disordered edges and will behave as quasi-ballistic for smooth disorder realizations. These results point towards a realistic implementation of valley polarized channels that can be achieved in current experimental settings by appropriate design of substrates or sample preparation. We propose two specific experimental scenarios that can implement the model presented here: 1) a stretched fold configuration may be produced starting with a longitudinal slit covered by a graphene membrane in a sealed container. By pressuring the container with a gas (helium for example) an out-of-plane fold-like structure  is formed with increased interatomic distance in the stretched region\cite{Bunch}; 2) another setup consists of a graphene membrane suspended on top of an extended longitudinal trench.  Out of plane stretching is achieved by pulling the suspended region of graphene by the gate voltage located at a small distance on top or at the bottom of the membrane to produce a fold-like deformation. In contrast to previous works that used similar configurations to study transport across the deformed region and predict vanishing of ballistic channels in the two-terminal conductance\cite{Fogler,Vozmediano2}, we show that transport {\it along} the stretched region in fact enhances ballistic transport in the direction parallel to the deformation. 

\section{Model}
The system is modeled by a zigzag terminated ribbon, with width and length $W, L$ on the vertical and horizontal directions respectively, and an extended out-of plane Gaussian deformation as shown in Fig.~\ref{Fig1} described by:
\be
h\left( y_{i}\right) =A
e^{-\frac{(y_{i}-y_{0})^2}{b^2}},
\label{gaussian}
\ee
with its center at $y_0=W/2$. $A$ and $b$ parametrize its amplitude and width, respectively. The strained fold axis is parallel to the ribbon length along the zigzag crystalline orientation. This particular geometry maximizes the effect of the deformation and produces optimal valley filtering as discussed below.

Electron dynamics is governed by a nearest neighbor tight-binding Hamiltonian 
\begin{equation}
H=\sum\limits_{<i,j>} t_{ij}c_i^\dagger c_j + h.c.\,\,,
\end{equation}
where, $c_i^\dagger$ ($c_i$) is the creation (annihilation) field operator in the $i$-th site, and $t_{ij}$ is the modified nearest-neighbor hopping energy $t_{ij} = t_{0}e^{-\beta \left(\frac{l_{ij}}{a}-1\right)}$. Here $t_{0}= -2.8 $eV, $a=1.42\AA$  (interatomic distance in unstrained graphene), and $\beta=\left|\frac{\partial\log t_o}{\partial\log a}\right| \simeq 3$. The deformation is described using elasticity theory\cite{Landau,Katsnelson} with strain tensor $\varepsilon_{\mu \nu}=\frac{1}{2}\left(\partial_\nu u_\mu+\partial_\mu u_\nu+\partial_\mu h \partial_\nu h\right)$, with the in- and out-plane deformation, $u_\nu$ and $h$, respectively\cite{Carrillo,Moldovan}. It is included in the model in the distance 
$l_{ij}=\frac{1}{a}\left(a^{2}+\varepsilon_{xx}x_{ij}^2+\varepsilon_{yy}y_{ij}^2+2\varepsilon_{xy}x_{ij}y_{ij}\right)$, where $x_{ij}$ and $y_{ij}$ correspond to the projected distance between sites $i$ and $j$ in $x$ and $y$ directions, before the deformation, respectively. According to the spatial dependence of the deformation (Eq.~\ref{gaussian}),  the new interatomic distances are given by
\begin{equation}
l_{ij}= a\left(1+\varepsilon_{yy} y_{ij}^{2}/a^{2}\right).
\label{interatomic-distance}
\end{equation}
reaching the maximum value $l_{ij}= a\left(1+\varepsilon_{yy}\right)$ for atomic positions separated by $(x_{ij}=0; y_{ij} = a)$.
This effective 2D model represents the change in hopping parameters due to changes in the overlap of $\pi$ orbitals that occur when the deformation forms ($\sigma$ bonds are not explicitly included). The effect of strain can be cast in terms of an inhomogeneous pseudo-gauge field $\vec{A}(\vec{r})$ with components 
\begin{equation}
\left(\begin{array}{c}
A_{x}\\
A_{y}
\end{array}\right)=\left(\begin{array}{c}
\varepsilon_{xx}-\varepsilon_{yy}\\
-2\varepsilon_{xy}
\end{array}\right)= \left(\begin{array}{c}
-2\frac{y^{2}}{b^{4}}h(y)^{2}\\
0
\end{array}\right)
\end{equation}
and pseudomagnetic field $\vec{B} = \bigtriangledown \times \vec{A}(\vec{r})$.

\begin{figure}
\includegraphics[scale=0.85]{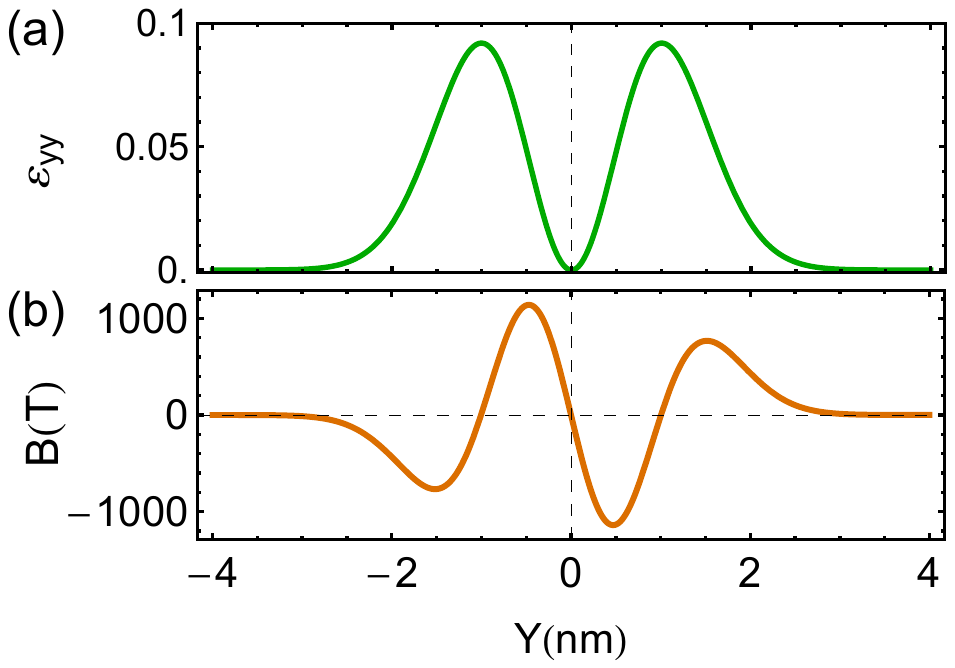}
\caption{(Color online) (a) Strain distribution and (b) pseudomagnetic field profile at $K$ valley due to an extended Gaussian out-of-plane deformation. Parameters: $W= 23.7 nm$, amplitude $A=0.7 nm$, and width $b=1.4 nm$.}
\label{Fig2} 
\end{figure}

From here on we use the parameter $\alpha = A/b$ to indicate the maximum strain intensity $\varepsilon_{m} = \alpha^{2}/e$ ($e = 2.71828...$), and to obtain the corresponding maximum pseudomagnetic field amplitude $B_{pm} \propto \varepsilon_{m}/b$. In Fig. \ref{Fig2} (a) we analyze the strain distribution across the transversal direction to the strained fold for fixed deformation parameters. In panel (b) the corresponding pseudomagentic field profile at valley $K$ is shown for the same fold parameters (the values of the pseudomagnetic field are reversed at valley $K'$). Notice, for example, that a value of $\alpha^{2}=25\% $ in Fig. \ref{Fig3} (a), corresponds to a maximum strain below $10\%$.

The conductance and LDOS are obtained with standard recursive Green's function techniques optimized for graphene systems\cite{Mucciolo}. To avoid spurious effects due to mode mismatching at the contacts, the strained fold is extended to the leads. When we consider disorder along the edges below, we do not include it in the leads.

Results below are for fixed size zigzag ribbons with the strained fold at its center. We verified that changes in the position of the strained fold center within a radius of $\sim 0.3 \text{nm}$ in the unit cell, as well as offsets in its position up to $4 \text{nm}$ with respect to the ribbon center for ribbons of different sizes ($W=8$ to $37 \text{nm}$), do not significantly modify conductance and LDOS properties.  In all cases studied the strained folds are fully embedded in the ribbon, i.e. in the regime $b/W \ll 1$.

\section{Conductance and LDOS} 

\begin{figure}
\includegraphics[scale=0.53]{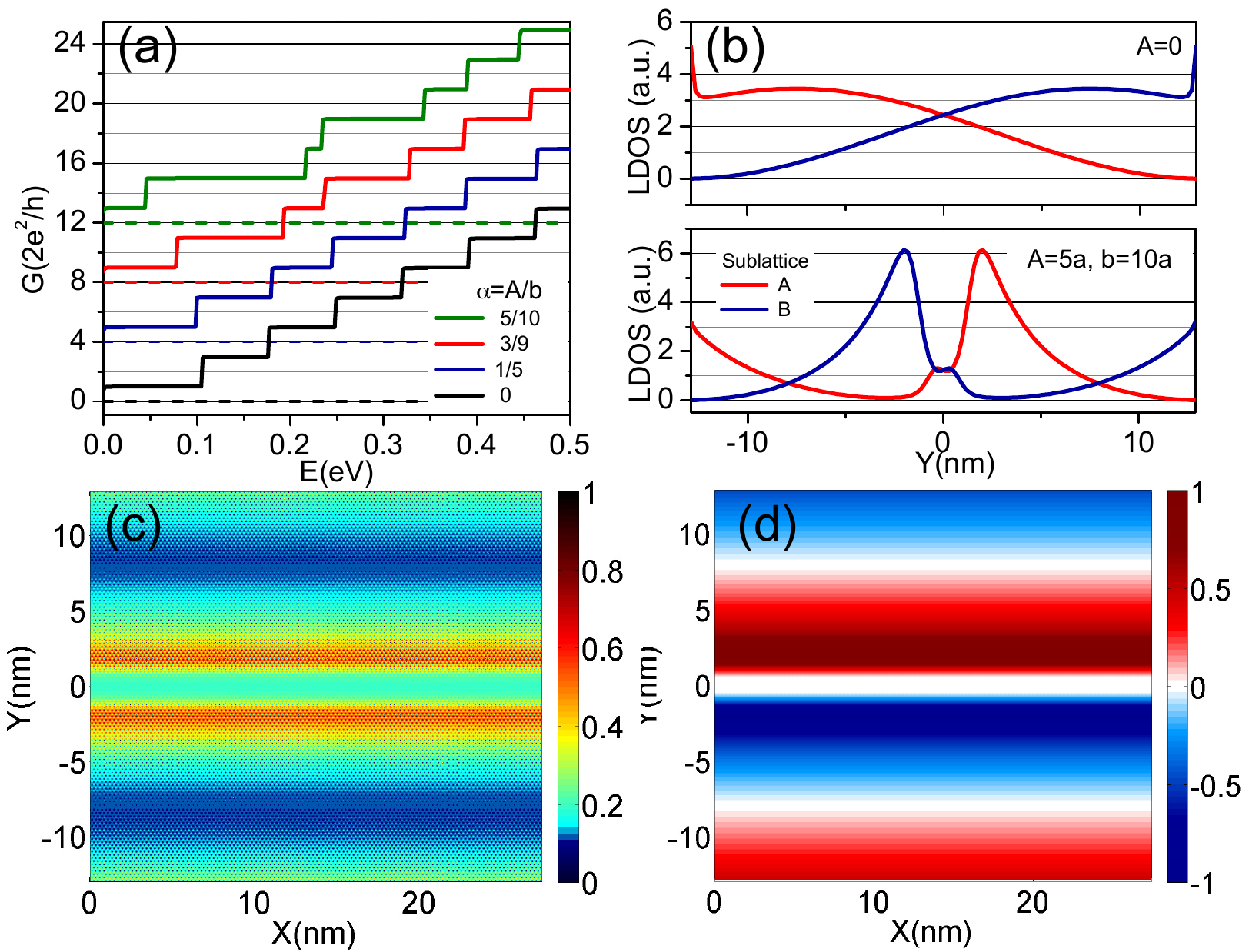}
\caption{(Color online) (a) Conductance for ribbon ($L = 27.4 nm$ and $W = 25.8 nm$) with different strained folds parameters. Curves are shifted for clarity. Dashed horizontal lines mark zero value for proper comparison. (b) LDOS profile across the ribbon with upper/lower panels  showing results for: $\varepsilon_{m} = 0\% - 9.2\%$ (black to green curves in (a)), (c) Enhanced total LDOS produced by the strained fold, and (d) Color map of sublattice polarization. Panels (b)-(d) obtained at $E=0.05 \text{eV}$. Parameters: $W=27.4 nm$, amplitude $A =  0.7 nm$, and width $b=1.4 nm$.
\label{Fig3}}
\end{figure}

Typical results for conductance are shown in Fig.~\ref{Fig2} (a) for no strain (black) and increasing strain values $\varepsilon_{m}= \alpha^{2}/e = 1.5\%$ (blue) to $9.2\%$ (green). For $\varepsilon_{m}=0$ (black) the first conductance plateau represents 2 ballistic channels (one per spin) due to edge states in the zero energy band while the second plateau contains 6 channels. As strain increases, the onset of the second conductance plateau moves to lower energies and becomes wider. The increase in width is produced by spectral transfer from other energies, and its onset  at lower energies represents an effective increase in the conductance. The number of channels contributing to the conductance within the energy range of the first plateau in the unstrained ribbon (energies below 0.1 eV in Fig.~\ref{Fig2} (a)) increases with strain from 2 to 6 (with 4 channels added to the existing 2). Notice that these changes are in contrast to those obtained in models with uniaxial in-plane strain that exhibit conductance gaps when transport occurs across the strained region\cite{Gradinar,Bahamon}. Panel (b) shows profiles of LDOS at energy $E = 0.05 eV$ across the ribbon, with the upper/lower panels showing results for ribbons with $\varepsilon_{m} =0$ (black) and $9.2\%$ (green). An enhanced LDOS develops around the deformed region with a similar spatial distribution to the exhibited by the pseudo-magnetic field (see Fig.\ref{Fig5} bottom of panels (a),(b)). The increase in LDOS at the edges corresponds to edge states. Panel (c) shows the total LDOS at $E = 0.05 eV$ along the ribbon, with enhanced values along the strained fold. Panel (d) exhibits the characteristic sublattice symmetry breaking that appears around the stretched area \cite{Carrillo,Martin}, as well as the one produced by perfectly terminated zigzag edges (upper and bottom sides). While this last one is due to the zigzag termination (A sites at one edge and B sites at the opposite edge), the former can be understood within the tight-binding model as due to the breaking of inversion symmetry in the unit cells of the underlying lattice. Once the strained regions is created, the center of inversion in each unit cell that characterizes pristine graphene is absent from those unit cells located at the sides of the strained fold axis, i.e, the corresponding lattice vectors measured from sites A and B to the fold axis are at different distances from it\cite{Sasaki2}. The data suggests that the extra conductance channels are composed by sublattice polarized states localized around the strained area.

\begin{figure}
\includegraphics[scale=0.25]{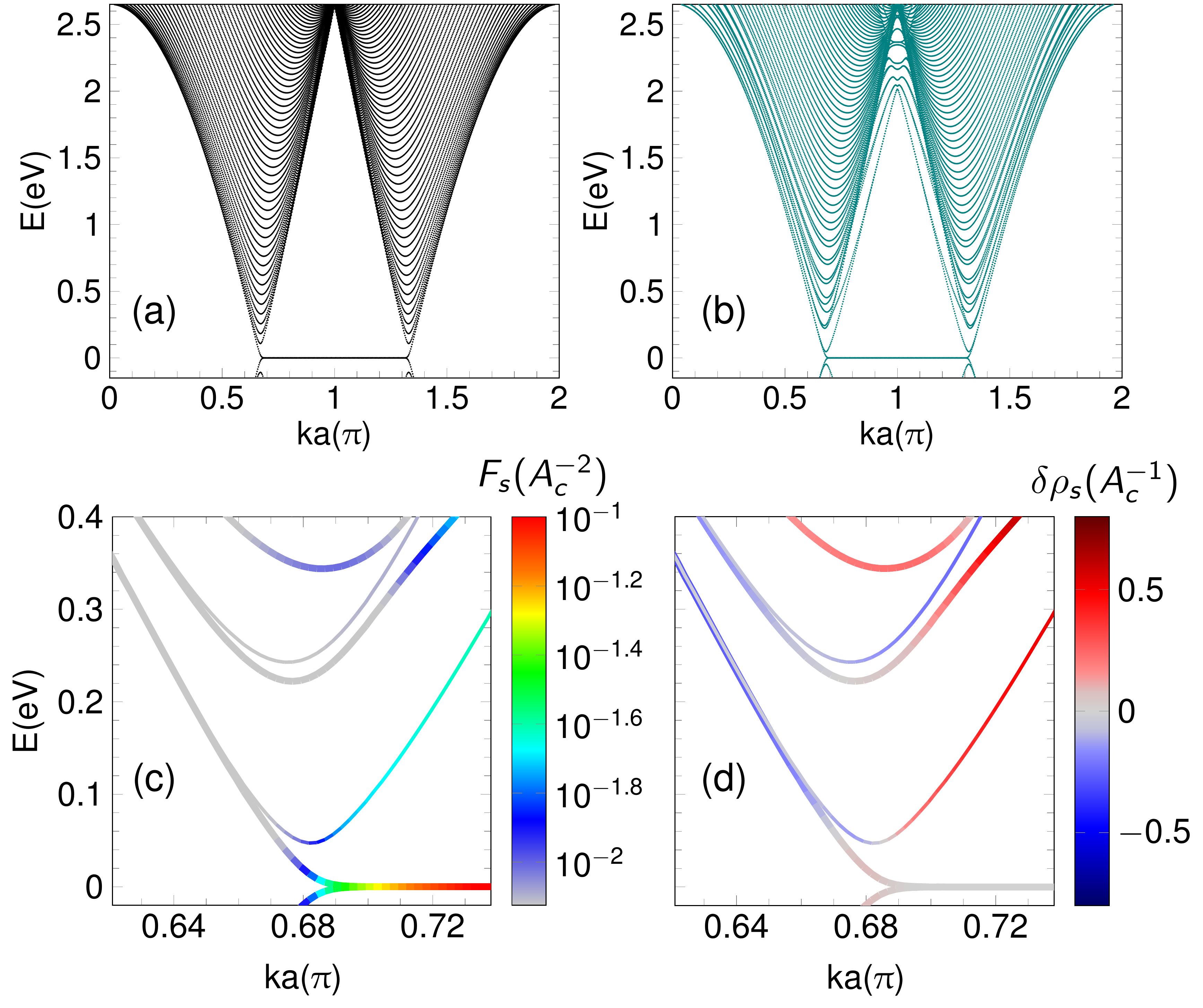}
\caption{(Color online) Band structure for a ribbon with a flat (a) and (b) a strained fold configurations. Panels (c) and (d) show a zoom in close to the Dirac point $K$ (c) Color scale indicates confinement level measured by parameter $F$ in units of $1/A_{c}^{2}$, ($A_{c}=$ unit cell area) as defined in main text. (d) Color scale indicates position across the ribbon measured by parameter $\delta \rho$ (in units of $1/A_{c}$) defined in main text. Parameters: $W= 23.7 nm$, amplitude $A=0.7 nm$, and width $b=1.4 nm$.} 
\label{Fig4}
\end{figure}

In order to further characterize these channels, we analyze the changes produced by strain in the band structure and wavefunctions. Figs.~\ref{Fig4} (a) and (b) show band structures for pristine and deformed ribbons, respectively. The effect of strain appears clearly at bands close to  zero-energy at the Dirac points as well as at higher energies near the band center. We confirmed that the positions of the $K, K'$ points shift towards each other with increasing strain as expected. Changes near the $K$ point are shown in panels $(c)$ and $(d)$. The color scale in panel $(c)$ represents values of the parameter $F$ (inverse participation ratio) that measures the degree of localization of states\cite{Ziman}. It is defined by $F = \sum_{i} |c_{i}|^{4}$, where $c_{i}$ represents the wavefunction amplitude at the $i$-th site\cite{ZhengBN} and the sum runs over all lattice sites.  For example, states near the center of the zero energy band are more localized than those near the Dirac point. The data show localization for states in higher energy bands in the presence of the deformation. In panel $(d)$ we introduce the parameter $\delta \rho = \sum_{i > i_{m}}\{ |c_{i}^{(\varepsilon)}|^{2} - |c_{i}^{(\varepsilon=0)}|^{2}\} $ to determine the real space position of these localized states. $\delta \rho$ is calculated adding contributions from sites $i$ around the strained region (with $i_{m}$ determined by $h(y_{i_{m}}) \ge 0.01 \AA$). The color code shows that these states belong to higher energy bands. Bands around $K'$ are mirror images of the ones shown here.
\begin{figure}
\includegraphics[scale=0.30]{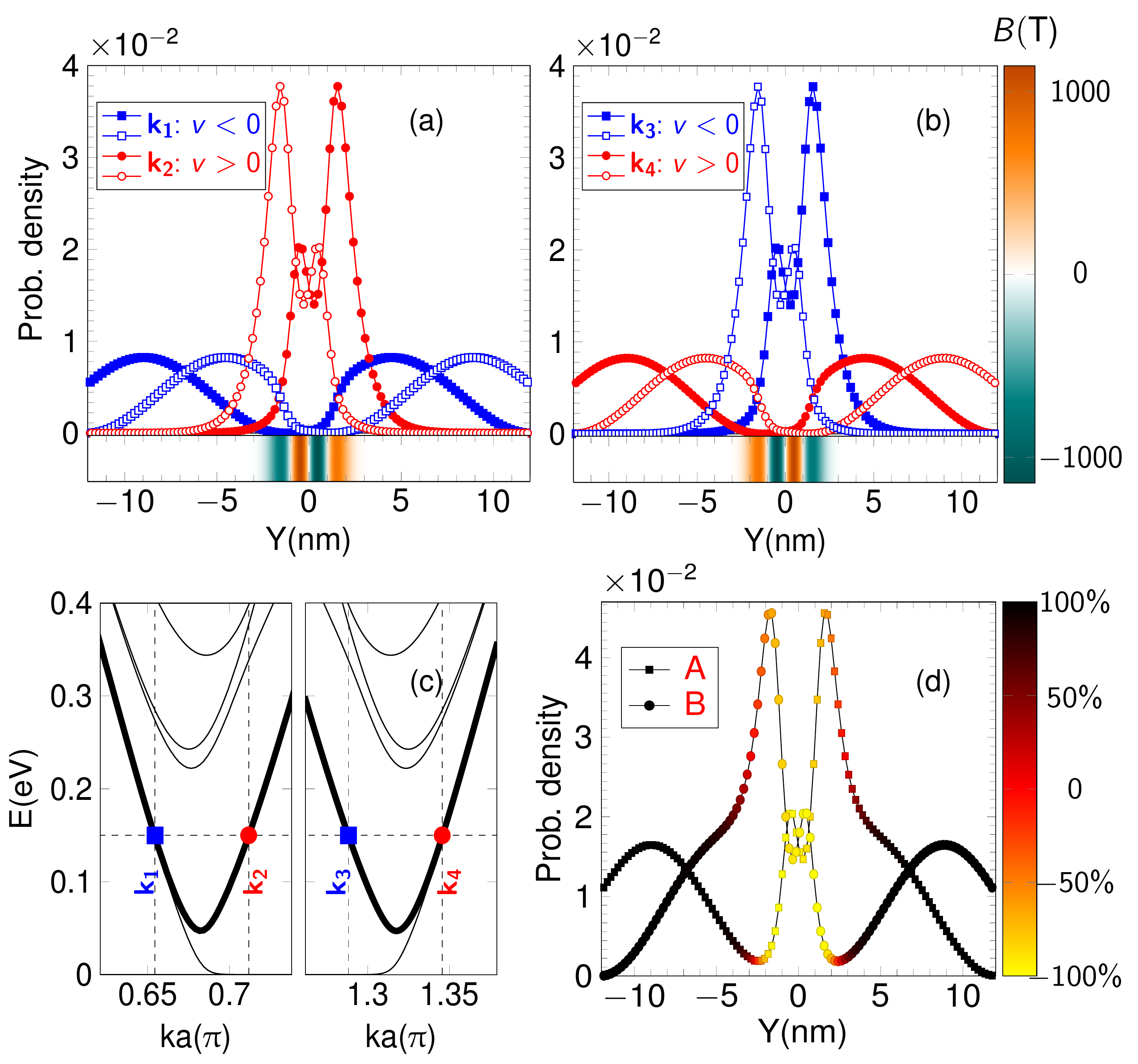}
\caption {(Color online) Probability densities for states at energy $E=0.15 \text{eV}$. Blue (red) curves correspond to state $k_1$ ($k_2$) with  negative (positive) velocity. Filled (empty) symbols indicate sublattice A (B). (a) States near Dirac point $K$. (b) States near Dirac point $K'$.  Color scale indicates magnitude of pseudo magnetic field (bottom). (c) Position of states in band-structure: right $K$ and left $K'$ respectively. (d) Total LDOS with states from both valleys with same velocity. Color scale indicates location in band structure: yellow near $K$ and black near $K'$. Parameters: $W=23.7 nm$, amplitude $A =  0.7 nm$, and width $b=1.4 nm$. 
\label{Fig5}}
\end{figure}

To visualize the real space distribution of these new localized states, we plot probability amplitudes in Fig.~\ref{Fig5} $(a)$ and $(b)$ for states at energy $E=0.15\text{ eV}$ located around $K$ and $K'$ respectively. The states, labeled by $k_1$ and $k_2$, are located at symmetric positions around both Dirac points (red and blue) as shown in panel $(c)$. Full and empty circles correspond to probability densities at sites in sublattices A and B, respectively. The color scale represents values of the pseudo-magnetic field at each valley, as depicted in the bottom part of the panels. Amplitudes of states with the same velocity ($v \propto \partial E /\partial k$), and originating at different valleys, appear larger in different regions across the strained fold. Thus, states from valley $K$ and positive velocity are concentrated at the center of the deformed region, while those from valley $K'$ have larger amplitudes along the sides of it. Panel $(d)$ shows a plot of the LDOS across the strained fold obtained by adding up all states at energy $E = 0.15 eV$ with the same velocity. The color code refers to states from valleys $K$ and $K'$, and identifies the valley separation: LDOS for states from valley $K$ is enhanced at the center (larger values of pseudomagnetic field) while that from valley $K'$ is larger at the sides. We have obtained similar results for different ribbon sizes and strain values. The numerical results confirm that the degree of valley filtering at fixed strain $\varepsilon_{m}$, is determined by the value of the pseudo-magnetic field, and it can be controlled by the ratio $b/W$. Narrow deformed structures exhibit better valley filtering properties with more focused conductance channels. 

\begin{figure}
\includegraphics[scale=0.42]{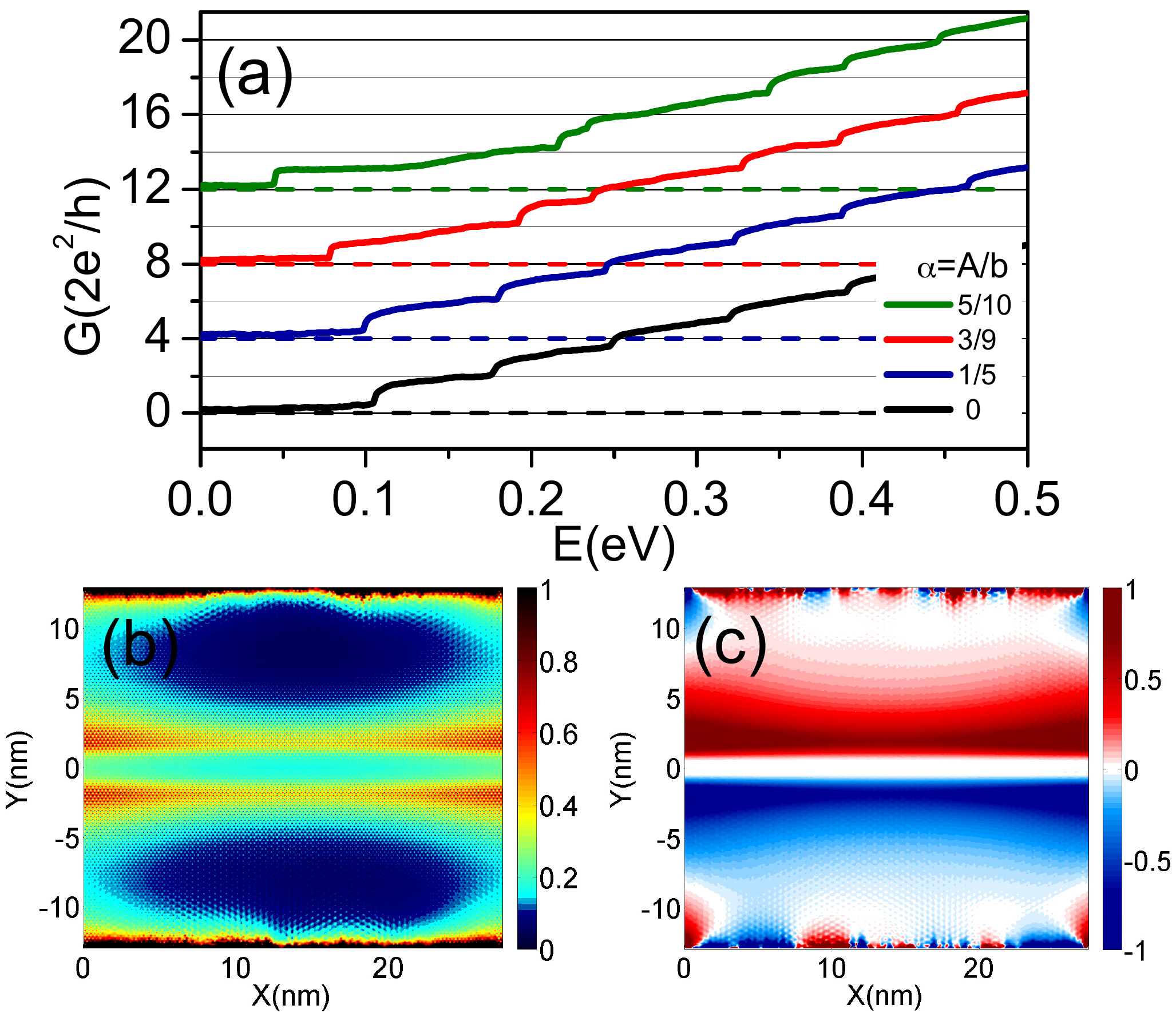}
\caption{(Color online) (a) Conductance for different fold amplitudes $A$ and widths $b$ averaged over 100 disorder realizations. Curves are  shifted for clarity purposes. Dashed horizontal lines mark zero value for proper comparison. (b) LDOS and (c) sub-lattice polarization averaged over 10 disorder realizations. Parameters: $A = 0.7\text{nm}$, $b = 1.4 \text{nm}$, and energy $E=0.05 \text{eV}$. Scales (a.u.) normalized to exhibit areas with higher density of states. (Black color regions along the edges correspond to highly disordered sites.)} 
\label{Fig6}
\end{figure}

In order to probe the stability of these states we include edge disorder in a finite central region using the implementation developed in Ref.~\onlinecite{Disorder}. In Fig.~\ref{Fig6} $(a)$ we show results for the conductance averaged over 100 disorder realizations. As reported \cite{Disorder}, the first conductance plateau is greatly diminished because disorder eliminates the contribution of edge states. However, there remains a well defined second conductance plateau that shifts to lower energies with increasing strain, similar to the clean ribbon case. The conductance is increased by only two ballistic channels because disorder eliminates edge channels (vanishing of first conductance plateau), and also those running along the sides of the strained area, with the other two remaining at the center, as shown in panels $(b)$ (LDOS), and $(c)$ (sublattice polarization). The enhanced LDOS region across the strained fold is reduced, and the sublattice polarization on the upper and lower parts of the figure is greatly diminished when compared to Fig.~\ref{Fig3}$(c)$ and $(d)$ (except at the edges where it is modulated by the disorder). It is important to note that edge disorder is the most important source of randomness in suspended samples\cite{APLreview}. As strained folds are suspended structures, defects inherited in the fabrication process are also a possible source of disorder, although unlikely in good quality samples that exhibit a minimum amount of vacancies or impurities that could cause short-range scattering. These features plus the spatial valley-separation between channels provides further protection against bulk disorder.

Finally, it is straightforward to show that strained folds with their principal axis at an angle $\theta$ with respect to the zigzag direction, produce fields with amplitudes $B_{\theta} = B_{zg} \cos(3\theta)$, where $B_{zg}$ is the pseudomagnetic field of a  strained fold along the zigzag direction. Thus, strained folds parallel to the armchair direction do not produce a field, and are not valley filters. These predictions can be tested by a proper alignment of the graphene membrane with respect to the direction of the trench on which it is suspended, or the substrate from which it is pulled from.

\section{Conclusions} In summary, we have studied the effects of strain created by an engineered strained fold on the transport and LDOS properties of a graphene ribbon. Our results show an enhanced LDOS around the deformed area with the expected sublattice symmetry breaking as reported for other out-of-plane deformations. Conductance calculations reveal extra channels within the energy range corresponding to the first conductance plateau for the undeformed ribbon, in addition to those due to edge states. Band structure calculations confirm that these channels originate from higher energy states that localize along the strained fold-like area. Furthermore, states with the same velocity show real space valley polarization, i.e., a current injected along the deformed structure will be split into two currents: one along the center of the strained fold constituted by states from one valley, and another running at its sides with contributions from states of the other valley as shown schematically in Fig.1. Disorder along the edges destroys the contribution from edge states, and from states localized farther away from the strained region center. Due to this spatial separation of states, the current is expected to be composed mostly by states from one valley at a given point. Different strained fold orientations will produce varying degrees of valley filtering with strained folds parallel to the zigzag direction being optimal valley polarizers. These findings can be tested in transport measurements in appropriately prepared substrates.

While finishing this manuscript we became aware of Ref.~[\onlinecite{Peeters3}] on valley filtering properties for armchair ribbons with a local out-of plane deformation designed to produce snake states\cite{Peetersold}, consistent with previous findings for local deformations\cite{Dawei}.

\noindent{\it Acknowledgments } We acknowledge
discussions with J. Mao, Y. Jang, D. Zhai, F. Mireles, M. Asmar and G. Petersen.
This work was supported by SBF-APS Brazil-USA (R.C.), NSF-DMR 1508325 (D.F., N.S.), FAPERJ E-26/101.522/2010
(A.L., D.F.), CNPq (A.L., C.L.), DOE-FG02-99ER45742, NSF-DMR 1207108 (E.A.).

\bibliography{Refs}
\end{document}